# Interaction of Nitrogen-Vacancy Centers in Diamond with a Dense Ensemble of Carbon-13

O.R. Rubinas[1,2,3], V.V. Soshenko[1,2], I.S. Cojocaru[1,2,3], S.V. Bolshedvorskii[1,2], P. G. Vilyuzhanina[3,4],

E.A. Primak[3,5], S.M. Drofa[3,5], A.M. Kozodaev[4], V.G. Vins[6], V.N. Sorokin[1,2], A.N. Smolyaninov[2] and

A.V. Akimov[1,2,3]

[1]*P.N. Lebedev Physical Institute RAS, Leninsky Prospekt 53, Moscow, 119991, Russia*

[2]*LLC Sensor Spin Technologies, Nobel St. 9, Moscow, 121205, Russia*

[3]*Russian Quantum Center, Bolshoy blvd., 30, p. 1, Skolkovo Innovation Center, Moscow, 121205, Russia*

[4]*National Research Nuclear University "MEPhI", 31, Kashirskoe Highway, Moscow, 115409 Russia*

[5]*Moscow Institute of Physics and Technology, 9 Institutskiy per., Dolgoprudny, Moscow Region, 141701, Russia*

[6]*LLC Velman, 1/3 st. Zelenaya Gorka, Novosibirsk ,630060, Russia*

email: a.akimov@rqc.ru

The nitrogen-vacancy center in diamond attracts a lot of attention in sensing applications, mainly for temperature, magnetic field, and rotation measurements. Nuclear spins of carbon-13 surrounding the nitrogen-vacancy center can be used as a memory or sensing element. In the current work, a diamond plate with a relatively large concentration of carbon-13 was synthesized and examined. The spectrum of optically detected magnetic resonance was recorded and analyzed in a magnetic field range of 5–200 G. A strain-independent measurement technique of carbon-13 isotope concentration based on the analysis of magnetic resonance spectra was developed. Additionally, narrow features in the spectrum were detected and understood.

# I. INTRODUCTION

Diamonds have long been a subject of fascination in material science due to their exceptional physical and optical properties, which make them valuable for a wide range of applications. In recent years, there has been a growing interest in the study of the interaction between nitrogen-vacancy (NV) centers [1] and the carbon-13 isotope ($^{13}$C) [2] in diamond, particularly in the context of quantum sensing [3] and information processing [4]. NV centers are particularly promising for these applications due to their long spin coherence times [5], high sensitivity to temperature [6], magnetic field [7,8], rotation [9,10], and the possibility of coherent control even at high temperatures [11]. The NV center consists of a nitrogen atom and a vacancy defect in the diamond lattice, which creates a localized electronic system [12] with an electronic spin equal to 1 that can be optically addressed and read out [13]. Meanwhile, $^{13}$C is a stable isotope of carbon with a nuclear spin of 1/2, making it an ideal candidate for quantum memory-assisted measurements [14,15], studying nuclear magnetic resonance [16,17], potential measurement of rotation [9], spin-imaging [18], and magnetic field sensors for chemical [19] and biological applications [20].

Interaction between NV centers and $^{13}$C ensembles in diamond have been considered previously [21,22]. Previous investigations have focused on exploring the interactions between a single NV center and small numbers of nearby $^{13}$C atoms. These interactions were examined separately in both theoretical and experimental studies [23,24]. Also, paper [22] demonstrated the optically detected magnetic resonance (ODMR) in diamond plates containing different concentrations of the $^{13}$C isotope, namely 1%, 10%, and 100%. Notably, the 10% $^{13}$C content predominantly resulted in NV centers with one closest $^{13}$C atom.

In the present investigation, a sample with an intermediate between 10% and 100% $^{13}$C content was considered. Here both one and two nearest $^{13}$C atoms are quite probable and influence the spectrum a lot. A detailed spectrum of all transition involved in ODMR was calculated for arbitrary concentration of $^{13}$C in diamond and compared to the experimental spectrum. In particular, the main microwave transitions contributing to the ODMR spectrum, were identified, including a transition weakly sensitive to the magnetic field, preserving a relatively narrow width even in the presence of a large number of paramagnetic impurities. Additionally, the behavior of the ODMR spectrum was analyzed and measured over a range of external magnetic field strengths from 5 to 200 Gauss. A method to measure the isotopic concentration of $^{13}$C, which is insensitive to strain and magnetic field effects, is proposed. The concentration measurements were verified by comparison with the Raman spectroscopy method and are congruent with the latter. However, in contrast with Raman spectroscopy, the suggested method is not sensitive to the strain of a diamond. In the case of Raman spectroscopy, changes in the concentration of $^{13}C$ and stress are indistinguishable. Thus, the suggested method has fewer systematics and allows for a more accurate determination of the isotope concentrations.

## II. DIAMOND PREPARATION.

Two samples under investigation were prepared via a high-pressure high-temperature diamond growth process in the Fe-Ni-C system. For the $^{13}C$-enriched sample ("enriched sample"), the carbon part of the growth system was composed of half graphite composed of a nearly pure isotope $^{12}C$ (99% purity) and half graphite composed of a nearly pure isotope $^{13}C$ (99% purity). NV centers were formed in the monocrystal due to post-growth electron irradiation with a power of 3 MeV and a dose of $1.5 \times 10^{18}$ cm$^{-2}$ in combination with subsequent vacuum annealing (1400 °C/2 hours) [25]. The second sample, namely the reference sample, was grown from graphite with natural $^{13}C$ content and treated the same way.

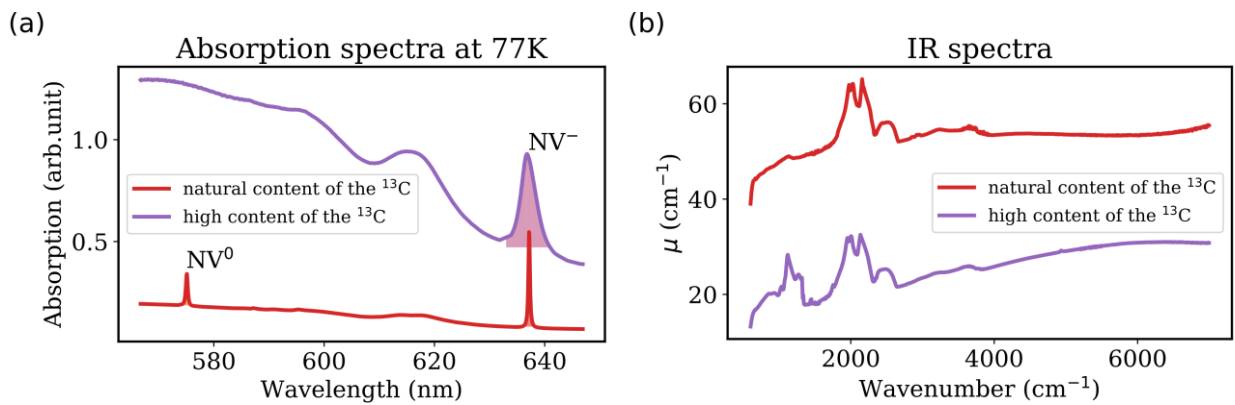

*Figure 1. Absorption spectra for enriched and reference samples: a) in the visible range at a temperature of 77 K; b) in the IR range at room temperature.*

To determine the NV center concentration in the treated samples, absorption and luminescence spectra were acquired using a Bruker Optik GmbH spectrometer at a temperature of 77 K (Figure 1A). It was found that the zero-phonon line (ZPL) at 637 nm for the enriched sample had a significantly larger width (FWHM = 3.4 nm) compared to that of the reference sample (FWHM = 0.3 nm) which was grown under similar conditions. This notable broadening of the ZPL is attributed to the high concentration of donor nitrogen in the sample under investigation [26]. No neutral-charged NV centers ($NV^0$) in the enriched sample were detected. The concentration of NV centers was determined using the integration of the ZPL intensity in the absorption spectrum technique [27]. It was determined that the concentration of NV centers in the enriched sample was

$11\pm 3$ ppm. The concentration of NV centers in the reference sample was measured as $1.5\pm 0.2$ ppm.

Also, the donor nitrogen ($p_1$ or C-centers) concentration was verified using the absorption coefficient in the IR spectrum (Figure 1B) as [28]:

$$n_{p1}[\text{ppm}] = 25\left[\frac{\text{ppm}}{\text{cm}^{-1}}\right] \times \mu\left[\text{cm}^{-1}\right], \qquad (1)$$

and was found to be $260\pm 20$ ppm in the enriched sample. The concentration of $p_1$ centers in the reference sample was significantly lower and estimated as $3-10$ ppm. Moreover, IR absorption at 1450 cm$^{-1}$, after irradiation and annealing of the sample corresponding to interstitial nitrogen $N_{int}$ [28], was observed in the diamond under study making it possible to measure the concentration of interstitial nitrogen to be $3\pm 1$ ppm.

### III. THE EXPERIMENT

ODMR in NV centers has been extensively studied [29] and is believed to be a result of the spin selective decay of the triplet excited state via singlet state (Figure 2A). This decay leads to a reduction in photoluminescence when the electron spin is polarized perpendicular to an axis of the color center. To measure the ODMR spectrum, an experimental setup as shown in Figure 2B was utilized. Diamond samples were mounted on a parabolic concentrator, which was connected to a long-pass filter and then a photodiode. The microwave antenna was realized as a coil [27] and placed directly around the diamond. Optical excitation of an ensemble of NV centers was achieved using a 532-nm laser (Coherent compass 315M), which was focused onto the sample with an objective (Olympus Plan 10X, NA=0.25).

The ODMR signal was obtained by the CW ODMR protocol [30]. During ODMR spectroscopy, the sample was exposed to the excitation laser light (Coherent compass 315M), thereby enriching the population of $m_S=0$ electron spin state. The laser power was 10 mW, and the beam diameter before the objective was 4.5 mm. The measurement procedure consisted of two parts, namely reference and signal measurement. During the first part, microwave irradiation was turned off, and diamond fluorescence was recorded forming the reference signal ($R$), corresponding to the polarized spin state. During the second part, microwave irradiation was turned on with power fed to the antenna of 2 mW. With microwave frequency equal to magnetic transition in the NV center, the population on $m_s = 0$ state decreases leading to a decrease in fluorescence. During the second phase, emission of the NV center ensemble was detected forming the ODMR signal ($S$).

The contrast of the observed ODMR resonance was defined as

$$C = \frac{S-R}{S+R} \times 100\% . \qquad (2)$$

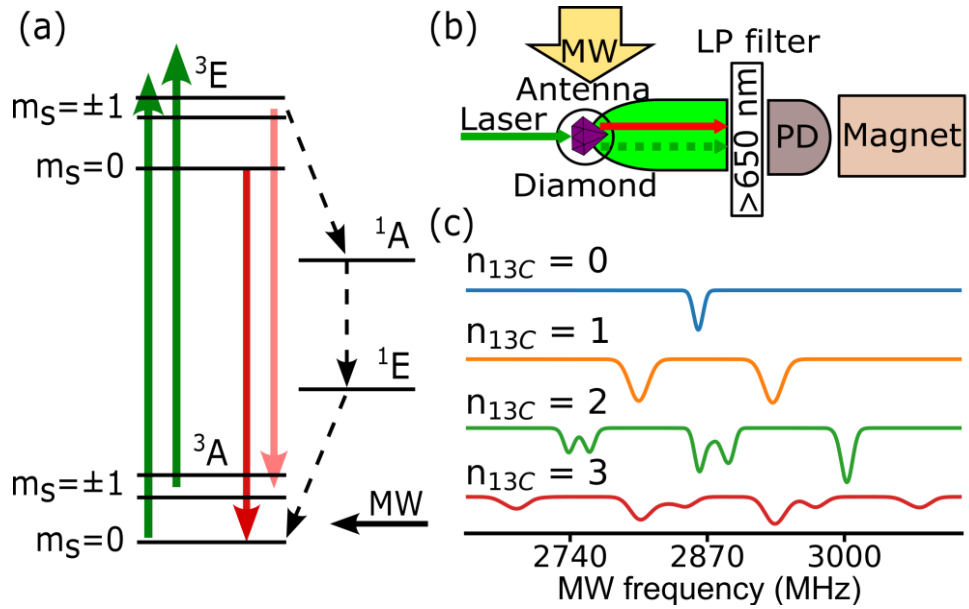

Figure 2. a) Energy-level system of the NV center; b) Setup; c) ODMR with different amounts $n_{13C} = 0,1,2,3$ of $^{13}C$ in the nearest cell in a zero magnetic field.

It has been previously demonstrated [29] that $^{13}C$ atoms located near the vacancy of the NV center have the most pronounced effect on the ODMR spectrum. The splitting of ODMR resonance caused by the interaction of the electron spin with the spin of nearby $^{13}C$ is 131 MHz, while in the absence of $^{13}C$, the ODMR line is centered on the zero-field splitting transition $D=2870$ MHz. In the case of a relatively large concentration of $^{13}C$, there is a significant probability of having 2 or even 3 $^{13}C$ atoms next to the vacancy, thus the ODMR spectrum of an ensemble should be the sum of 4 contributions, with no $^{13}C$ next to vacancy, with one, with two, and with all three carbons being $^{13}C$, as indicated at Figure 2C. The spectrum is further complicated by splitting each of the resonances caused by the $^{13}C$ spin bath located further from the vacancy, as well as the interaction with the nitrogen nuclear spin of $^{14}N$, leading to a significant increase in the number of observed lines around the ones indicated in Figure 2C. The splitting caused by the nuclear spin of nitrogen is well studied and is around 2.2 MHz. The spectrum is also affected by the strain in the sample and magnetic field.

The contributions to the ODMR spectrum from NV centers with different numbers of $^{13}C$ closest to the NV center ($n_{13C}$) may be inferred from the overall ODMR spectrum in a nearly 0 magnetic field. Figure 3A has two side doublets (marked as 2), corresponding to the case of $n_{13C} = 2$, two strong lines corresponding to $n_{13C} = 1$ (marked as 1), and a doublet structure in the center of ODMR spectra corresponding to the combination of $n_{13C} = 2$ and $n_{13C} = 0$ contributions (marked as 2 and 0, respectively). Thus, it is expected that the case of $n_{13C} = 3$ is not significant and thus sets a limitation on the concentration of $^{13}C$ in the sample.

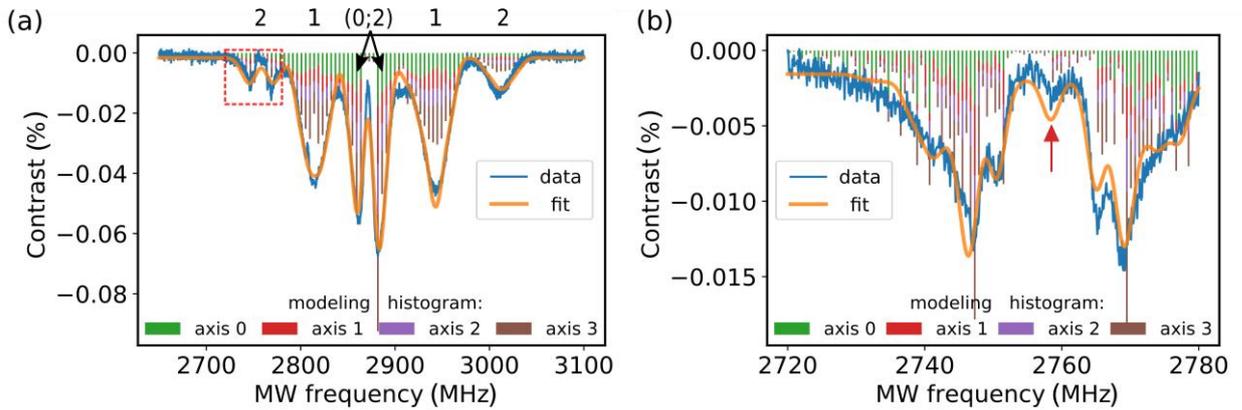

*Figure 3. a) The ODMR spectrum obtained for $^{13}C$-enriched diamond in a 5 G magnetic field; b) Zoomed region of interest in the spectrum marked by the red dashed rectangle in (a), with the red arrow indicating the location of the transition with low magnetic sensitivity. The orange curve is a fit function (Appendix I), and the histogram represents all acceptable transitions in the NV-$^{13}C$ system (Appendix II). The histogram is color-coded to represent the four subgroups of NV centers oriented along the four possible directions of the crystal lattice. The numbers 0, 1, and 2 denote the resonances that arise due to the interaction with the corresponding number of nearby $^{13}C$.*

## IV. DETERMINATION OF THE CARBON-13 CONCENTRATION IN DIAMOND

The diamond lattice allows for 4 possible orientations of NV centers in the case of monocrystal diamond. To understand some of the parameters, it is convenient to isolate resonance from one specific orientation of the NV center. Therefore, to determine the concentration of $^{13}C$, a magnetic field of 200 Gauss was applied (Figure 4), enabling the isolation of one of the orientations, as presented in Figure 5A. To achieve this, a magnet was installed in a manner such that the field direction coincided, within the error, with the orientation of one of the NV centers as the source of the field. The magnitude and direction of the magnetic field were determined using a reference sample with a natural content of $^{13}C$ (Figure 4A). Both the enriched sample and reference sample, each measuring approximately 1×1×0.5 mm in size, were placed on the same parabolic concentrator placed within the ring antenna. Switching between them was accomplished by shifting the pump beam, thereby enabling ODMR measurements in the same magnetic field.

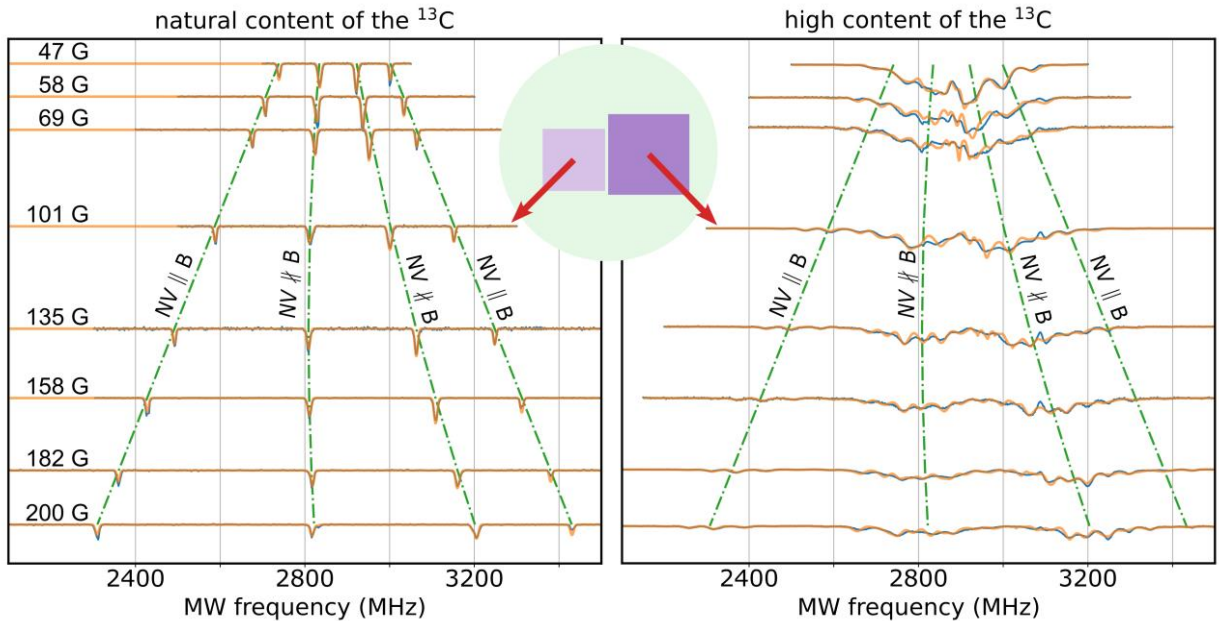

*Figure 4. NV centers ensemble ODMR spectra for various magnetic fields: for the reference sample with a natural abundance of $^{13}$C on the left; for the sample with 30% of $^{13}$C on the right. In the center – a schematic of two diamonds placed on the same parabolic concentrator. The dashed green line shows the numerical calculation of the resonant frequency for $N=0$ in different magnetic fields.*

The fundamental concept behind the concentration measurements lies in the 4-component nature of the spectrum, resulting from splitting due to nearby $^{13}$C atoms. If the probability to find $^{13}$C in the selected node of the diamond lattice is $p$, one could easily calculate probabilities $P$ to have 0,1,2,3 $^{13}$C next to the vacancy of the NV center as follows:

$$\begin{aligned} P_0(p) &= (1-p)^3 \\ P_1(p) &= 3p(1-p)^2 \\ P_2(p) &= 3p^2(1-p) \\ P_3(p) &= p^3 \end{aligned} \quad , \qquad (3)$$

where the index stands for the number of $^{13}$C next to the vacancy.

Hence, the relative amplitudes of the spectral components at characteristic frequencies contain information about the likelihood of finding $^{13}$C in a specific position and consequently, they provide information about the relative concentration of $^{13}$C out of all carbon atoms:

$$\frac{n_{13C}}{n_C} = p \qquad (4)$$

Accordingly, the concentration of $^{13}C$ could be extracted from the relative amplitudes of major features of the spectrum. To understand the width, one needs to calculate the entire spectrum of all 271 lines (see Supplementary data). Undoubtedly, the width of the resonances also depends on the concentration of $^{13}C$ via the splitting of resonances, depending on the distance to particular $^{13}C$, as well as dephasing, caused by the random distribution and the dynamics of surrounding spins. Nevertheless, the amplitude of the resonance has a very clean dependence on $p$, and therefore the spectrum could be fitted using generalized width $\sigma$ for all $k$ transitions. The transition strengths $F_{ik}$ are calculated according to equations (12) – (17) (see section VI) for each weight $P_i(p)$. Each transition is approximated with a normalized Gauss shape. Thus, the fit function $F(f)$ for the entire spectrum is:

$$F(f) = b + A \sum_{i=1}^{4} \sum_{k} P_i(p) F_{ik}^{2} \frac{1}{\sqrt{\pi}\sigma} e^{-(f-f_{ik})^2/\sigma^2} \qquad (5)$$

The model was then used to fit the spectrum of selected resonance and extract the relative concentration of $^{13}C$ to be equal to $p = 27.6 \pm 0.1\%$. Here $b$ is the signal offset, $A$ is the total scale factor, $f$ is frequency, $f_{ik}$ are transition frequencies calculated also from the full Hamiltonian of the system (see section VI). The fit parameters were only concentration $p$, width $\sigma$, total scale factor $A$, and the projection of the magnetic field $B_{NV}$ on the chosen NV orientation. The initial guess for the magnetic field $B = 200$ G was estimated from the reference sample as shown in Figure 4. The value of the field was selected to be able to isolate the single NV orientation in the ODMR spectrum.

As an alternative to the elaborated model, a simpler approach with 7 Gaussian lines was utilized to define carbon isotope concentration. Taking into account that the amplitude of the resonance decreases with $n_{13C}$ as $2^{-(n_{13C}+1)}$, the following formula was utilized:

$$F'(f) = b + A \sum_{n=0}^{2} \sum_{k=0}^{2^n} \frac{P_n(p)}{2^{n+1}} \frac{1}{\sqrt{\pi}\sigma} e^{-(f-f_{nk})^2/\sigma^2} \qquad (6)$$

The spectrum (Figure 5a) has only 5 resonances visible while the model uses 7. To address this issue, frequencies were fixed for ambiguous resonances, namely $f_{21} = f_{00} + 2, f_{22} = f_{00} + 31$ (shown with arrows in Figure 5a), where 2 and 31 MHz shift from $f_{00}$ frequency is obtained from the elaborated model. The width of Gaussian contours is set to be equal for all resonances as the main widening mechanism is the same for all, namely the spin bath. Fitting with this model results in a slightly lower result of $p = 27.1 \pm 0.1\%$. The concentration error is calculated solely from the fit error covariance matrix. Therefore, a simpler fitting model with only 7 Gaussians provides a good estimation of concentration, deviating by 0.5% from the more complex ODMR spectrum fitting model.

The key factor that could lead to systematic errors in concentration is the frequency dependence of the microwave field amplitude, as it will distort the ODMR spectrum and change the resonance amplitude ratio. The microwave loop that was used is considered to be non-resonant and should not affect concentration estimation. The amplitude dependence was estimated from the datasheet for the amplifier and switch, used in the setup. With a peak-to-peak amplitude fluctuation of 5% within the ODMR frequency range, the estimated absolute change in concentration was 0.1%. Overall, the estimated concentration levels obtained through two different methods, suggested above, differ by approximately 0.5%. Considering that the sum of all uncertainties approaches a similar value, the uncertainty in the concentration measurements is estimated to be 0.5%.

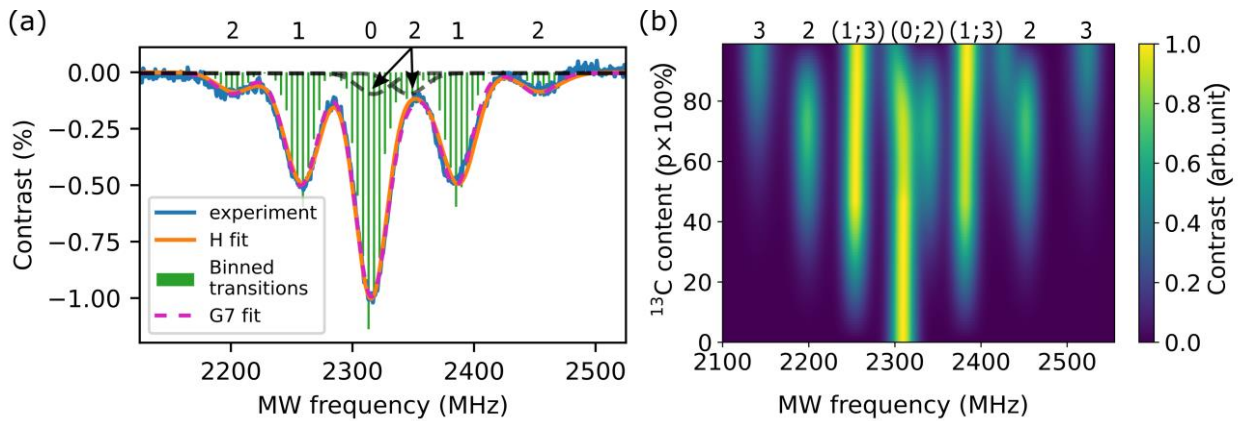

*Figure 5. a) ODMR spectrum for determining the concentration $^{13}$C in the enriched sample. The green histogram shows all possible transitions with weights corresponding to the strength. The orange line shows the fit function with the complete model (5); the black dashed line shows fit with the model (6); the black dash-dotted line shows "hidden" Gaussian lines with $f_{21}$, $f_{22}$ frequencies; the green bars show the histogram of transitions, the blue solid line presents the experimental data. b) Numerical calculations of ODMR hyperfine splitting due to different contents of $^{13}$C at 200 Gauss for the co-axis orientation of the NV center.*

The result of this assessment can be compared with the classical measurement of isotope concentration using Raman spectroscopy. To find the fraction of $^{13}$C isotope, one could use the following equation for the Raman diamond line shift $\nu$ [31]:

$$\left(1332.8\,\text{cm}^{-1} - \nu\right) - 34.77\,\text{cm}^{-1} p - 16.98\,\text{cm}^{-1} p^2 = 0, \tag{7}$$

where $\nu$ is a Raman shift. This formula, nevertheless, does not account for strain in diamond. The strain P in the sample contributes to the Raman shift in the following way [32,33]:

$$\Delta\nu = \frac{P}{0.34\,\text{GPa}}\,\text{cm}^{-1}, \tag{8}$$

The strain in diamonds can be independently estimated via the shift of the ZPL in the absorption spectrum [34]. As is described in section II, the line shifts were $1.9\pm0.3$ and $2.6\pm0.3$ meV for the reference and enriched samples, respectively. Using the coefficient 5.75 meV/GPa from this work [34], the strain pressure can be estimated as $0.33\pm0.05$ and $0.45\pm0.05$ GPa for the reference and enriched samples, respectively. It should be noted that the effect of the isotopic shift on the ZPL position is unknown. Nevertheless, if these values are used, one can calculate shifts of the Raman line from (8), it would correspond to 1.0 cm$^{-1}$ for the reference sample and 1.3 cm$^{-1}$ for the enriched sample.

The diamond line position was estimated for both plates from Raman spectra (see Figure 6). The spectra were approximated with the Lorentzian shape. The diamond scattering line wavenumber was found to be 1332.1 cm$^{-1}$ and 1320.3 cm$^{-1}$ for the reference and enriched samples, respectively. The used Raman spectrometer InVia Basis (Renishaw) exhibits excellent spectrum-to-spectrum stability of 0.01 cm$^{-1}$; however, to mitigate systematic errors, a reference sample with a known line is typically used. By utilizing the reference sample with the known natural $^{13}$C concentration ($p = 0.011$ [35]) and measured strain, the systematic shift was estimated to be $0.4\pm0.1$ cm$^{-1}$. Using formula (7) and considering the strain shift (8), the $^{13}$C concentration for the enriched sample was estimated to be $27\pm2\%$, with the primary source of error being the estimation of strain in the diamonds used. The result of concentration estimation is congruent with the suggested above method.

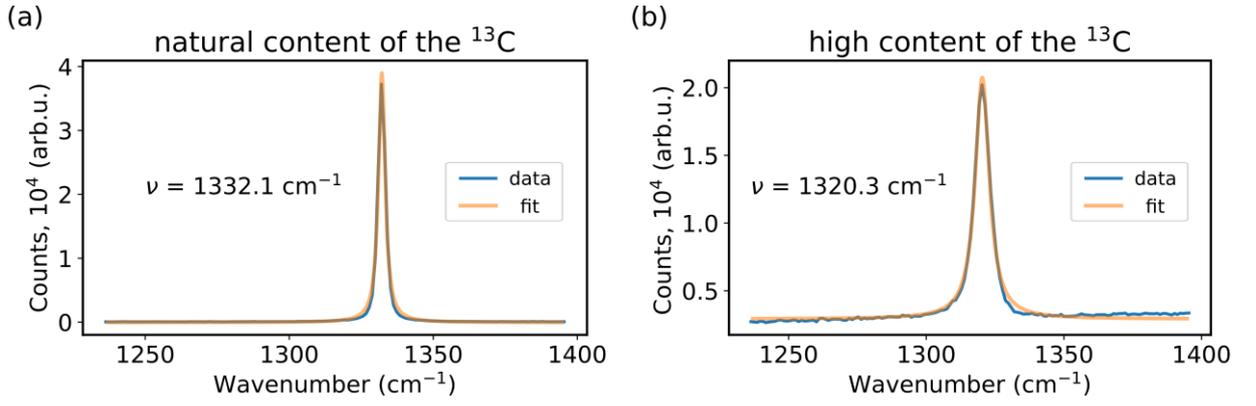

*Figure 6. Raman spectrum for a) reference sample; b) $^{13}C$-enriched sample.*

## V. DETAILS OF THE ODMR SPECTRUM

A deeper understanding of the spectrum requires calculation of the spectrum. Following the ideas of [21], the transitions of all possible transitions were calculated (see section VI). Since the concentration of $^{13}C$ and the strain were already determined, these parameters were fixed during simulations.

The observed width of the resonances in the spectrum is attributed to the interaction between the NV center and a $^{13}C$ spin bath. In this study, a model with a fixed width of lines was utilized. Since the magnitude of the hyperfine interaction varies with distance, only the closest atoms were considered. The interaction with the spin bath leads to the emergence of additional transitions that broaden the fundamental resonance. A spin bath formation model is also provided in APPENDIX I. Besides, the electron spin bath also contributes to the broadening of the resonances in the spectrum. According to [36], the broadening effect caused solely by 300 ppm of donor nitrogen can be estimated empirically as 10 MHz FWHM.

Figure 3 presents the fit of the experimental data with the model developed. It can be seen that while there are some minor discrepancies, the overall fit describes the experimentally observed spectrum rather well. The parameters used in the fitting procedure are: the magnetic field $B$ and two angles between it and (111) diamond direction in a spherical axis, two angles of MW polarization, strain components $E_x$ and $E_y$, ZFS with the z-component of strain $D' = D + E_z$, $^{13}C$ content $p$, the width of transition lines $\sigma$, the spectrum level and scale. The transition width is sensitive to the surrounding spin bath, the width $\sigma$ was expressed as $\sigma = \sigma_0 + \Delta m_S \sigma_B$ where $\sigma_0$ is the width of transition with $\Delta m_S = 0$, and $\sigma_B$ is the parameter describing the broadening of the transitions due to the spin bath. Since the spectrum contains both sensitive and weakly sensitive to the magnetic field transitions, change in the electron spin projection $\Delta m_S$ is different for different transitions. To visualize the transition, taken into account in the simulation, a histogram displaying all possible transitions and the relative amplitude is presented.

Figure 3B demonstrates the zoom of the ODMR signal into the range of $2720-2780$ MHz. In this range, one can notice a narrow feature, the width of which is smaller than that of surrounding resonances (9). This is due to the fact that this transition can be written down in the basis with the defined components of electron and nuclear spins on the NV center axis as:

$$0.704|m_S = 0, m_I = 0, \uparrow, \downarrow\rangle + (-0.703 - 0.0287i)|m_S = 0, m_I = 0, \uparrow, \downarrow\rangle \rightarrow$$
$$\rightarrow 0.697|m_S = +1, m_I = 0, \downarrow, \downarrow\rangle + 0.697|m_S = -1, m_I = 0, \uparrow, \uparrow\rangle \quad (9)$$

$$0.706|m_S = 0, m_I = 0, \uparrow, \uparrow\rangle + (0.353 - 0.611i)|m_S = 0, m_I = 0, \downarrow, \downarrow\rangle \rightarrow$$
$$\rightarrow 0.697|m_S = +1, m_I = 0, \downarrow, \downarrow\rangle + 0.697|m_S = -1, m_I = 0, \uparrow, \uparrow\rangle \quad (10)$$

Here $m_S$ stands for the component of the electron spin of the NV center on its axis, $m_I$ stands for the nuclear spin of the nitrogen atom, included into the NV center, the arrows $\uparrow$ and $\downarrow$ stand for the orientation of two $^{13}$C located right next to the NV center. Due to the complete symmetry of magnetic moments involved, this transition turns out to be practically insensitive to the external magnetic field (Figure 7a), thus is noticeably narrower than other resonances, and has a Lorenz shape. In comparison, the magnetically sensitive transition:

$$0.998|m_S = 0, m_I = 0, \downarrow\rangle \rightarrow 0.992|m_S = 1, m_I = 0, \downarrow\rangle \quad (11)$$

depends linearly on a longitudinal magnetic field.

Unfortunately, the numerical coefficients in (9) are the function of the magnetic field, and already at the field ~5 G, the excited state of this transition becomes $|m_S = 1, m_I = 0, \downarrow\downarrow\rangle$, thus losing the symmetry. Still, in the range of low magnetic fields, this transition is somewhat different from others in its width. For this study, this resonance was approximated at 2760 MHz with a 3 MHz FWHM and approximated by the Lorentz contour. It should be noted that according to the simulations, there are a few resonances like this (10), but most of those seem to have very low amplitude and tend to lose the superposition state even at lower fields. Therefore, these resonances are not treated as special cases.

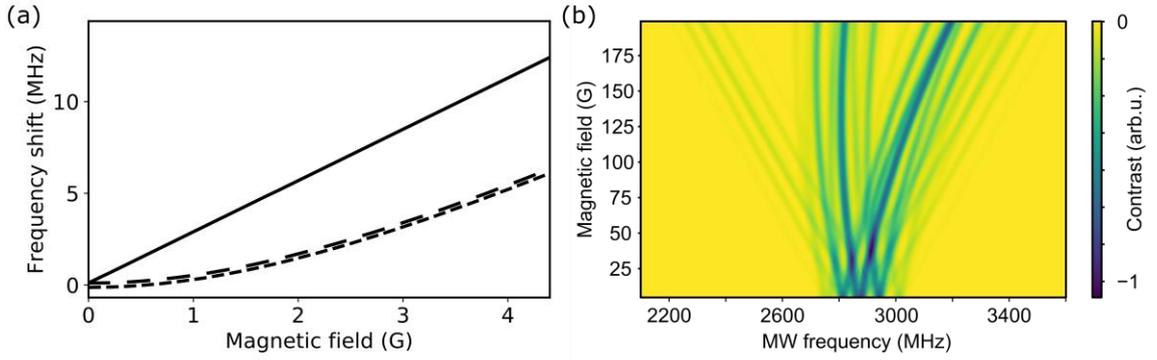

*Figure 7. (a) The dependence of the ODMR shift on the magnetic field relative to $B = 0$. The legend presents $n_{13C}$ for the corresponding plot and transitions (9) – long dash black line, (10) – short dash black line and (11) – solid blue line. (b) Numerical calculations of ODMR spectra in various external magnetic fields.*

Simulations that allow predicting the spectrum at the external magnetic field are presented in Figure 7b and summarized in Table 1, as well as Supplementary Data. Remarkably, due to the large complexity of the spectrum, 4 different orientations of the NV center can be separated into 4 isolated groups only at a field of 200 G.

The experimental data, presented in Figure 4, were fitted as follows: First, the data of the reference sample was analyzed using the fit function of the theoretical model with the number of $^{13}$C next to the NV center set to 0. The error in determining the field can be estimated from the width of the resonance on the reference sample and is approximately 0.06 Gauss. The experimental data for the enriched sample (Figure 4b) were approximated by the same fit function but with all possible combinations of $^{13}$C, excluding 3 simultaneous $^{13}$C next to the NV center, since the trace of those was not observed in the spectra.

## VI. THEORETICAL MODEL

The NV center possesses an electron spin with a value of $S = 1$ and a nuclear spin of $I = 1$ [12]. The energy level system of the ground state of the NV center can be characterized by the spin Hamiltonian, which comprises several fundamental interactions: the zero field splitting (ZFS) $H_0$; the Zeeman splitting $H_Z$ in the magnetic field $\vec{B}$; the strain interaction term $H_S$; quadrupole interactions $H_S$ with the nuclear spin of nitrogen $\vec{I}$, the hyperfine interaction within the NV center $H_{hf1}$, and interaction with the surrounding nuclear spin of $^{13}$C $H_{hf2}$. The overall Hamiltonian $H_{gs}$ thus could be written as:

$$
\begin{aligned}
H_{gs} &= H_0 + H_Z + H_S + H_Q + H_{hf1} + H_{hf2} \\
H_0 &= DS_z^2 \\
H_Z &= \gamma_e \vec{B}\cdot\vec{S} - \gamma_n \vec{B}\cdot\vec{I} \\
H_S &= E_z S_z^2 + E_x(S_y^2 - S_x^2) + E_y(S_x S_y + S_y S_x) \\
H_Q &= Q I_z^2 \\
H_{hf1} &= A_{N\|} S_z I_z + A_{N\perp}(S_x I_x + S_y I_y)
\end{aligned}
\quad (12)
$$

Here $\vec{S}$ is the spin of electrons in the NV center with a magnitude of 1, $D = 2870$ MHz [13], $\gamma_e = 2.8$ MHz/gauss and $\gamma_n = 1.1\cdot 10^{-3}$ MHz/gauss are the gyromagnetic ratios for electron and nuclear spins, respectively [36], $E_i$ are components of the strain field; $A_{N\|} = -2.15$ MHz, $A_{N\perp} = -2.6$ MHz are the axial and transverse hyperfine constants, respectively, and $Q = -4.95$ MHz is the nuclear electric quadrupole parameter [36].

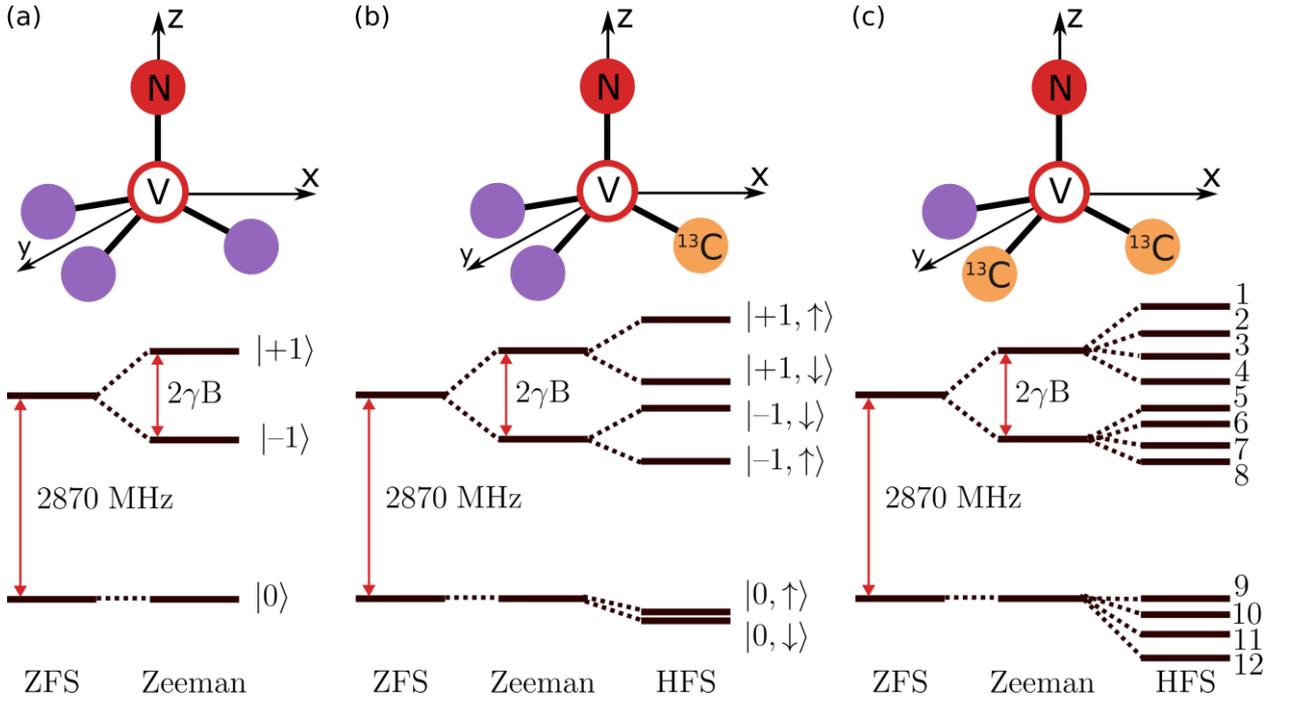

Figure 8. NV center in the diamond lattice and its ground state in an external magnetic field ≈100 Gauss: A) without $^{13}C$ in the nearest shell; B) with one $^{13}C$; C) with two.

| Number in Figure 1C | E, MHz | State |
|---|---|---|
| 1 | -19.13 | $0.7\|0,\uparrow\downarrow\rangle + 0.7\|0,\downarrow\uparrow\rangle$ |

| 2 | -11.63 | $0.7\|0,\uparrow\uparrow\rangle - (0.35 - 0.6i)\|0,\downarrow\downarrow\rangle$ |
| 3 | -7.50 | $0.7\|0,\uparrow\uparrow\rangle + (0.35 - 0.6i)\|0,\downarrow\downarrow\rangle$ |
| 4 | -0.73 | $0.7\|0,\uparrow\downarrow\rangle - 0.7\|0,\downarrow\uparrow\rangle$ |
| 5 | 2737.10 | $0.7\|1,\downarrow\downarrow\rangle - 0.7\|-1,\uparrow\uparrow\rangle$ |
| 6 | 2754.59 | $0.7\|1,\downarrow\downarrow\rangle + 0.7\|-1,\uparrow\uparrow\rangle$ |
| 7 | 2846.34 | $0.5\|1,\uparrow\downarrow\rangle - (0.5 - 0.2i)\|1,\downarrow\uparrow\rangle + (0.35 - 0.36i)\|-1,\uparrow\downarrow\rangle - (0.17 - 0.46i)\|-1,\downarrow\uparrow\rangle$ |
| 8 | 2853.60 | $0.5\|1,\uparrow\downarrow\rangle + (0.45 - 0.19i)\|1,\downarrow\uparrow\rangle + (0.34 - 0.35i)\|-1,\uparrow\downarrow\rangle + (0.19 - 0.47i)\|-1,\downarrow\uparrow\rangle$ |
| 9 | 2895.23 | $0.5\|1,\uparrow\downarrow\rangle - (0.47 + 0.16i)\|1,\downarrow\uparrow\rangle - (0.35 - 0.35i)\|-1,\uparrow\downarrow\rangle + (0.44 - 0.22i)\|-1,\downarrow\uparrow\rangle$ |
| 10 | 2905.93 | $0.5\|1,\uparrow\downarrow\rangle + (0.47 + 0.17i)\|1,\downarrow\uparrow\rangle - (0.34 - 0.36i)\|-1,\uparrow\downarrow\rangle - (0.47 - 0.23i)\|-1,\downarrow\uparrow\rangle$ |
| 11 | 3003.85 | $0.7\|1,\uparrow\uparrow\rangle + (0.27 + 0.65i)\|-1,\downarrow\downarrow\rangle$ |
| 12 | 3005.13 | $0.7\|1,\uparrow\uparrow\rangle - (0.27 + 0.65i)\|-1,\downarrow\downarrow\rangle$ |

*Table 1. Numerically calculated eigen energies and eigenstates of the ground state Hamiltonian for the NV center with two $^{13}C$ atoms in the nearest shell in a magnetic field $B = 100$ G, the numbers 1–12 correspond to Figure 8c levels.*

The hyperfine splitting of the sublevels associated with $^{13}C$ in the NV center $H_{hf2}$ is significantly stronger than that of nitrogen [36], exhibiting a difference of two orders of magnitude [2]. However, in the calculations and simulations presented in this article, the contribution of nitrogen was also taken into account. Depending on the amount of $^{13}C$ in the system, the hyperfine interaction Hamiltonian can be expressed in different ways [24,37]:

$$H_{hf2}^{(1)} = \sum_{i,j} A_{13Ci,j}^{(1)} S_i \otimes I_j^C,$$

$$H_{hf2}^{(2)} = H_{hf2}^{(1)} \otimes \begin{pmatrix} 1 & 0 \\ 0 & 1 \end{pmatrix} + \sum_{i,j} A_{13Ci,j}^{(2)} S_i \otimes \begin{pmatrix} 1 & 0 \\ 0 & 1 \end{pmatrix} \otimes I_j^C,$$

(13)

where $I^C$ is the nuclei spin of $^{13}C$ (with a magnitude of 1/2), $A_{13C}^{(1)}$ and $A_{13C}^{(2)}$ are the hyperfine coupling tensors for $N=1$ and $N=2$ numbers of $^{13}C$ in the nearest cell.

The hyperfine interaction tensor $A_{13C}$ is described at [24] as:

$$A_{13C} = \begin{pmatrix} 123.3 & & \\ & 123.3 & \\ & & 204.9 \end{pmatrix} \quad (14)$$

The matrix $A_{13C}$ is given in the coordinate system of the NV center. However, as stated in the paper [24], it is necessary to consider the transition $R^{(1)}$ to a unified coordinate system due to the geometric arrangement of the NV center and $^{13}C$ in diamond. Consequently, the hyperfine interaction tensor with $^{13}C$ is represented as:

$$A_{13C}^{(1)} = R^{(1)} A_{13} R^{(1)T}$$
$$R^{(1)} = \begin{pmatrix} 1 & 0 & 0 \\ 0 & -0.2742 & 0.9617 \\ 0 & -0.9617 & -0.2742 \end{pmatrix} \quad (15)$$

In the case where two or more $^{13}C$ isotopes are present in the system, the transition matrix $R^{(2)}$ was calculated by rotating $R^{(1)}$ by 120° around the z-axis (co-axial to the axis of NV centers).

To calculate all the energy levels in the ground state of all NV center groups in the ensemble, the Schrödinger equation was solved for the Hamiltonians formed for different $n_{13C}$. A numerical solution was obtained using the QuTip package in Python. It is worth noting that the theoretical description of the ensemble of NV centers requires finding a solution for each orientation, as the magnetic field projection on each of the four directions of the diamond lattice will differ, affecting the values of the Zeeman contribution to the Hamiltonian.

The Hamiltonian for the NV center (1) + microwave field system is as follows [36]:

$$H = H_{gs} + H_{MW} = H_{gs} - \gamma_e \vec{S} \vec{b}_{MW} \cos \omega_{MW} t, \quad (16)$$

where $\vec{b}_{MW}$ and $\omega_{MW}$ are the oscillating microwave field amplitude and frequency, correspondently, $t$ is time. For further evaluations, it is necessary to move to the basis $U$ of the $H_{gs}$ operator eigenstates:

$$\begin{aligned} H' &= H'_{gs} + H_{MW}' \\ H'_{gs} &= U^{\dagger} H_{gs} U \\ H'_{MW} &= U^{\dagger} H_{MW} U \end{aligned} \quad (17)$$

The obtained operator $H'$ characterizes the transitions between the NV-center states. Next, the rotating wave approximation [36] was applied to the matrix elements $h_{ij}$ of the operator $H'$ responsible for the microwave field-resonant transitions. In essence, this means neglecting the rapidly oscillating terms that appear in the interaction Hamiltonian and retaining only those that oscillate at a much lower rate close to resonance. Thus, the transition strength $F_{ij}$ can be estimated by Rabi frequency as $F_{ij} \approx h_{ij}/2^2 \approx \Omega_R^2$. Given the population dynamics, the transition amplitude $A$ is more strictly equal to:

$$A \propto 1 - \cos \frac{\pi}{2} \left( \frac{\Omega_R \alpha}{1 + \Omega_R \alpha} \right) \quad (18)$$

where $\alpha$ is the ratio of microwave pumping and decay rates during a set of relaxation processes. Applying the theoretical model to a real experiment, the polarization direction of the microwave antenna was taken into account. Therefore, the Hamiltonian of interaction with the microwave field was written as:

$$H_{MW} = -\gamma_e b_{MW} \left( S_y \sin \varphi_{MW} + S_z \cos \varphi_{MW} \right) \cos \omega_{MW} t \ , \quad (19)$$

where $\varphi_{MW}$ is the angle between microwave antenna polarization and the NV-center axis.

The above model was used to describe and approximate the ODMR spectra. Specifically, equation (12) was used to calculate the values of eigenstates and transitions between them. The contribution of each transition to the spectrum was estimated as the product of (18) and (3). By knowing in advance the value and direction of the external magnetic field $\vec{B}$, the strain in the diamond $\vec{E}$, the $^{13}C$ concentration in the diamond $p$, and the orientation of the microwave antenna polarization in space, the resonant frequencies $\omega_0$ and the amplitudes $A$ of the ODMR spectrum can be calculated. In this work, these parameters were used as approximation parameters to construct the fit function. The constant of ZFS $D$ was also taken as an approximation parameter since it can depend on the temperature [38] and the $\alpha$ parameter from equation (18). To approximate the resonant peaks, a Gaussian contour with some standard deviation $\sigma$ was used:

$$G = A \frac{1}{\sigma\sqrt{2\pi}} e^{-\frac{1}{2}\left(\frac{\omega-\omega_0}{\sigma}\right)^2} \quad (20)$$

where $A$ is the transition amplitude, and $\omega_0$ is resonant frequency. The standard deviation of the Gaussian contour was expressed as $\sigma = \sigma_0 + \Delta m_S \sigma_B$, where $\sigma_0$ corresponds to magneto-insensitive transitions and $\sigma_B$ to magneto-sensitive ones. This standard deviation was taken as an approximation parameter and related to the transverse relaxation time through the expression $\sigma = 1/\sqrt{2}\pi T_2^*$. To determine the orientation of the magnetic field in space, it was referenced to a given coordinate system shown in [39] using two angles in spherical coordinates: $\theta$ for the zenith angle and $\varphi$ for the azimuthal angle. Additionally, two angles ($\xi, \zeta$) were introduced to specify the polarization orientation of the microwave antenna in space. The fit function also included the spectrum level and overall scaling factor as parameters (see APPENDIX II).

## VII. CONCLUSION

In this study, the sample with a large concentration of $^{13}C$ was studied in detail. The method for determining the concentration of the $^{13}C$ isotope of carbon based on the analysis of the relative amplitude of resonances was proposed and used to determine the concentration of $^{13}C$ in the sample to be $p = 27.6 \pm 0.5\%$. The method is not sensitive to the intrinsic strain of diamond. The results of measurements were compared with measurements based on Raman spectroscopy and made it possible to confirm the contents of $^{13}C$. The strain of the diamond plate, affecting Raman measurements, was estimated from the ZPL shift. The spectrum of ODMR was recorded and analyzed in the range magnetic field of 5–200 G. All key transitions in the spectrum were identified using the developed theoretical model.

## VIII. ACKNOWLEDGMENTS

This research was supported by grant No. 21-42-04407 of the Russian Science Foundation.

## APPENDIX I.  SPIN BATH MODELING

To accurately estimate the $^{13}$C spin bath influence on the NV center spectrum, carbon-13 atoms were considered in 9 sites closest to the vacancy, besides the 3 sites, previously described in this work. Hyperfine interaction for carbon-13 atoms in these sites was taken from [37] and is $A_{zz}^{(1)} = 13.7$ MHz, $A_{zz}^{(2)} = 12.8$ MHz for the first group of 6 and the second group of 3 atoms, respectively. Only the longitudinal part of the interaction was taken into account. Spin bath widen spectra were obtained by considering all possible configurations of $^{13}$C atoms within the mentioned 9 sites and calculating allowed transition frequencies and strength. The resulting ODMR spectrum for a certain $^{13}$C concentration $p$ was obtained as the weighted average of all configurations where weight corresponds to the probability of configuration given by binomial distribution.

## APPENDIX II.  FITTING DETAILS

The fit function $\Phi(Ex, Ey, D, B, \theta, \varphi, \xi, \zeta, \alpha, level, scale, \sigma_0, \sigma_B, p)$ calculates the frequencies and amplitudes of transitions in the system of NV centers and $^{13}$C under study and constructs an ODMR spectrum consisting of the sum of Gaussian contours at these transitions based on the specified approximation parameters. Some of the approximation parameters defined in one part of the work are used as fixed values in other parts (Supplementary Materials). To perform the fitting, the curve_fit method from the Python SciPy library was utilized, which minimizes the functional of $\Phi$ with respect to the parameters using classical optimization techniques.